\documentclass[12pt,epsf,epsfig,psfig]{article}
\usepackage{graphicx}
\usepackage{epstopdf}
\usepackage{epsfig}
\usepackage{cite}
\def\q{q{\bar q}}
\def\Q{Q{\bar Q}}

\def\E{\epsilon^{\rm vac}}

\def\tx{\tilde x}
\def\tt{\tilde t}

\def\be{\begin{equation}}
\def\ee{\end{equation}}

\def\lsim{\raise0.3ex\hbox{$<$\kern-0.75em\raise-1.1ex\hbox{$\sim$}}}
\def\gsim{\raise0.3ex\hbox{$>$\kern-0.75em\raise-1.1ex\hbox{$\sim$}}}

\def\NP{{ Nucl.\ Phys.\ }}
\def\PL{{ Phys.\ Lett.\ }}
\def\PR{{ Phys.\ Rev.\ }}

\def\PRL{{ Phys.\ Rev.\ Lett.\ }}

\def\ZP{{ Z.\ Phys.\ }}
\def\EP{{ Europ.\ Phys.\ J.\ C}}

\begin{document}

rev.\ 10.\ 6.\ 07 
~\hfill BNL-NT-07/18

\hfill BI-TP 2007/06

\vskip2.5cm

\centerline{\Large \bf Thermal Hadronization and}

\bigskip

\centerline{\Large \bf Hawking-Unruh Radiation in QCD}

\vskip2cm 

\centerline{\large \bf P.\ Castorina$^{a}$, D.\ Kharzeev$^{b}$ and 
H.\ Satz$^{c}$} 

\bigskip

\centerline{a) Dipartimento di Fisica, Universit{\`a} di Catania,}

\centerline{and INFN Sezione di Catania, Via 
Santa Sofia 64 I-95123 Catania, Italia}

\centerline{b) Physics Department, Brookhaven National Laboratory, 
Upton, NY 11973-5000, USA}

\centerline{c) Fakult\"at f\"ur Physik, Universit\"at Bielefeld, 
D-33501 Bielefeld, Germany}



\vskip2cm

\centerline{\large \bf Abstract}

\bigskip

We conjecture that because of color confinement, the physical vacuum forms 
an event horizon for quarks and gluons which can be crossed only by quantum 
tunneling, i.e., through the QCD counterpart of Hawking radiation by black 
holes. Since such radiation cannot transmit information to the outside, it 
must be thermal, of a temperature determined by the chromodynamic force at 
the confinement surface, and it must maintain color neutrality. We explore 
the possibility that the resulting process provides a common mecha\-nism for 
thermal hadron production in high energy interactions, from $e^+e^-$ 
annihilation to heavy ion collisions. 

\vfill
Keywords: \\
event horizon, color confinement, Hawking-Unruh radiation, 
multihadron production\\
\\
PACS: 04.70Dy, 12.38Aw, 12.38Mh, 12.40Ee, 25.75Nq, 97.60Lf
\\

\newpage

\section{Introduction}

The aim of this paper is to develop a conceptual framework for 
a universal form of thermal multihadron production in high energy 
collisions. Our work is based on two seemingly disjoint observations:
\begin{itemize}
\vspace*{-0.1cm}
\item{Color confinement in QCD does not allow colored constituents to
exist in the physi\-cal vacuum, and thus in some sense creates a situation
similar to the gravitational confinement provided by black holes.}
\vspace*{-0.2cm}
\item{Numerous high energy collision experiments have provided strong evidence 
for the thermal nature of multihadron production, indicating a universal
hadronization temperature $T_H\simeq 150-200$ MeV.}
\vspace*{-0.1cm}
\end{itemize}
We want to suggest that quantum tunnelling through a color event horizon,
as QCD counterpart of Hawking-Unruh radiation from black holes, can relate 
these observations in a quite natural way.

\medskip

The idea that the color confinement of quarks and gluons in hadrons may 
have a dual description in terms of a theory in curved space--time is not 
new. Both gravitational confinement of matter inside a black hole \cite{C-R} 
and the de Sitter solution of the Einstein equations with a cosmological 
constant describing a ``closed'' universe of constant curvature \cite{Salam} 
have been proposed as possible descriptions of quark confinement. Soon it 
became clear that asymptotic freedom \cite{Gross:1973id} and the scale 
anomaly \cite{scale1,scale2} in QCD completely 
determine the structure of low--energy gluodynamics \cite{MSh}. This 
effective theory can be conveniently formulated in terms of the 
Einstein--Hilbert action in a curved background. At shorter distances 
(inside hadrons), the effective action has the form of classical 
Yang--Mills theory in a curved (but conformally flat) metric \cite{KLT-Pom}. 
The ``cosmological constant'' present in this theory corresponds to the 
non--perturbative energy density of the vacuum, or the ``gluon condensate''  
\cite{SVZ}. 

\medskip

It is worthwhile to also mention here the well--known conjectured holographic
correspondence between the large $N$ limit of supersymmetric Yang--Mills 
theory in $3+1$ dimensions and supergravity in an anti--de Sitter space--time 
sphere, $AdS_5 \times S_5$ \cite{Maldacena:1997re}. This example illustrates 
the  possible deep relation between Yang--Mills theories and gravity; 
however, a conformal theory clearly differs from the examples noted  
above, in which the scale anomaly (describing the breaking of conformal 
invariance by quantum effects) was used as a guiding principle for 
constructing an effective curved space--time description.    

\medskip

Let us assume that color confinement indeed allows a dual description in 
terms of the gravitational confinement of matter inside black holes. What 
are the implications of this hypothesis for hadronic physics? Hawking 
\cite{Hawking} showed that black holes emit thermal radiation due to quantum 
tunneling through the event horizon. Shortly afterwards, Unruh \cite{Unruh}
demonstrated that the presence of an event horizon in accelerating frames also 
leads to thermal radiation. It was soon conjectured that the periodic 
motion of quarks in a confining potential \cite{Grillo}, or the acceleration 
which accompanies inelastic hadronic collisions \cite{Barshay,Hosoya,Horibe}, 
are associated with an effective temperature for hadron emission. 

\medskip

Recently, 
a QCD--based picture of thermal production based on the parton description of 
high energy hadronic collisions has been proposed \cite{K-T,KLT}. The 
effective temperature $T$ is in this case determined either by the string 
tension $\sigma$, with a relation 
\be
T= \sqrt{{3 \sigma \over 4\pi}},
\label{thagedorn}
\ee
or, in the gluon saturation regime, by the saturation momentum $Q_s$ 
describing the strength of the color fields in the colliding hadrons 
or nuclei, with $T  \simeq (Q_s/2\pi)$.  

\medskip

Turning now to the second observation, we recall that over the years, hadron 
production studies in a variety of high energy collision experiments have 
shown a remarkably universal feature. From 
$e^+e^-$ annihilation to $p-p$ and $p-\bar p$ interactions and further 
to collisions of heavy nuclei, with energies from a few GeV up to the 
TeV range, the production pattern always shows striking thermal aspects, 
connected to an apparently quite universal temperature around
$T_H \simeq 150 - 200$ MeV \cite{Hagedorn}. As a specific illustration 
we recall that the relative abundance of two hadron species $a$ and $b$, 
of masses $m_a$ and $m_b$, respectively, is essentially determined by the 
ratio of their Boltzmann factors \cite{species},
\be
R(a/b) \sim \exp-\{(m_a-m_b)/T_H\}.
\ee
What is the origin of this thermal behaviour? While high energy heavy ion 
collisions involve large numbers of incident partons and thus could allow 
invoking some ``thermalisation'' scheme through rescattering, in $e^+e^-$ 
annihilation the predominant initial state is one energetic $q \bar q$ pair, 
and the number of hadronic secondaries per unit rapidity is too small to 
consider statistical averages. The case in $p-p/p-\bar p$ collisions is
similar. 

\medskip

This enigma has led to the idea that all such collision experiments 
result in the formation of a strong color field ``disturbing'' the 
physical vacuum. The disturbed vacuum then recovers by producing hadrons 
according to a maximum entropy principle: the actually observed final state 
is that with the largest phase space volume. While this provides an 
intuitive basis for a statistical description, it does not account for 
a universal temperature. Why don't more energetic collisions result in
a higher hadronization temperature?

\medskip

A further piece in this puzzle is the observation that the value of the
temperature determined in the mentioned collision studies is quite similar 
to the confinement/deconfinement transition temperature found in lattice 
studies of strong interaction thermodynamics \cite{T-c}. While hadronization 
in high energy collisions deals with a dynamical situation, the ener\-gy loss 
of fast color charges ``traversing'' the physical vacuum, lattice QCD 
addresses the equilibrium thermodynamics of unbound vs.\ bound color
charges. Why should the resulting critical temperatures be similar or even
identical?  

\medskip

We shall here consider these phenomena as reflections of the QCD counterpart
of the Hawking radiation emitted by black holes \cite{Hawking}. These 
ultimate stellar states provide a gravitational form of confinement and
hence, as already noted, their physics was quite soon compared to that of 
color confinement
in QCD \cite{C-R,Salam}, where colored constituents are confined to 
``white holes'' (colorless from the outside, but colored inside). It
should be emphasized from the outset that in contrast to the original
black hole physics in gravitation, where confinement is on a classical 
level complete, in QCD confinement refers only to color-carrying 
constituents; thus, e.g., photons or leptons are not affected.

\medskip

In black hole physics, as noted above, it was shown that the event
horizon for systems undergoing uniform acceleration leads to quantum
tunnelling and hence to thermal radiation \cite{Unruh}. Our aim here
is to show that such Hawking-Unruh radiation, as obtained in the 
specific situation of QCD, provides a viable account for the thermal
behavior observed in multihadron production by high energy collisions.
Furthermore, in the process we also want to elucidate a bit the common
origin of the ``limiting temperature'' concepts which have arisen in 
strong interaction physics over the years.

\medskip

We begin by reviewing those features of black hole physics and Hawking 
radiation which are relevant for our considerations, and then discuss how 
they can be implemented in QCD. In particular, we show that modifications
of the effective space-time structure, in a perturbative approach as well
as in a non-perturbative treatment based on a large-scale dilaton field,
lead to an event horizon in QCD.

\medskip

Following this, we present the main conceptual consequences of our conjecture.
\begin{itemize}
\vspace*{-0.1cm}
\item{Color confinement and the instability of the physical vacuum under 
pair production form an event horizon for quarks, allowing a transition 
only through quantum tunnelling; this leads to thermal radiation of a 
temperature $T_Q$ determined by the string tension.}
\vspace*{-0.2cm}
\item{Hadron production in high energy collisions occurs through a
succession of such tunnelling processes. The resulting cascade is a
realization of the same partition process which leads to a limiting
temperature in the statistical bootstrap and dual resonance models.}
\vspace*{-0.2cm}
\item{The temperature $T_Q$ of QCD Hawking-Unruh radiation can depend only 
on the baryon number and the angular momentum of the deconfined system.
The former could provide a dependence of $T_Q$ on the baryon number density, 
while the angular momentum pattern of the radiation allows a 
centrality-dependence of $T_Q$ and elliptic flow.} 
\vspace*{-0.2cm}
\item{In kinetic thermalization, the initial state information is
successively lost through collisions, converging to a time-independent  
equilibrium state. In contrast, the stochastic QCD Hawking radiation 
is ``born in equilibrium'', since quantum tunnelling {\sl a priori} 
does not allow information transfer.} 
\vspace*{-0.2cm}
\end{itemize}

\vskip0.5cm

\section{Event Horizons in Gravitation and in QCD}

\subsection{Black holes}

A black hole is formed as the final stage of a neutron star after 
gravitational collapse \cite{Ruf}. It has a mass $M$ concentrated in 
such a small volume that the resulting gravitational field confines 
all matter and even photons to remain inside the event horizon $R$ of 
the system: no causal connection with the outside is possible. As a 
consequence, black holes have three (and only three) observable properties: 
mass $M$, charge $Q$ and angular momentum $J$. This section will address 
mainly black holes with $Q=J=0$; we shall come back to the more general 
properties in section 4. We use units of $\hbar = c = 1$.  

\medskip

The event horizon appears in a study of the gravitational metric, which
in flat space has the form
\be
ds^2 = g~\!dt^2 - g^{-1}~\!dr^2 - r^2[d\theta^2 + \sin^2\theta d\phi^2],
\label{metric}
\ee
using units where $c=1$. The field strength of the interaction is contained 
in the coefficient $g(r)$,
\be
g(r) =\left(1 - {2GM\over r}\right),
\ee
leading back to the Minkowski metric in the large distance limit 
$r\to \infty$. The vanishing of $g(r)$ specifies the Schwarzschild 
radius $R$ as event horizon,
\be
R= 2~\!G~\!M. 
\ee
It is interesting to note that the mass of a black 
hole thus grows linearly with $R$, analogous to the behaviour of the
confining potential in strong interactions: $M(R) = (2G)^{-1}\ R$.

\medskip

Classically, a black hole would persist forever and remain forever
invisible. On a quantum level, however, its constituents (photons,
leptons and hadrons) have a non-vanishing chance to escape by tunnelling 
through the barrier presented by the event horizon. Equivalently, 
we can say that the strong force field at the surface of the black
hole can bring vacuum fluctuations on-shell. The resulting Hawking 
radiation \cite{Hawking} cannot convey any information about the internal 
state of the black hole; it must be therefore be thermal. For a 
non-rotating black hole of vanishing charge (denoted as Schwarzschild black 
hole), the first law of thermodynamics,
\be
dM = T dS
\ee
combined with the area law for the black hole entropy \cite{Bekenstein},
\be
S = {\pi R^2 \over G}
\ee
leads to the corresponding radiation temperature 
\be
T_{BH} = {1 \over 8 \pi~\! G~\! M}.
\label{T-BH}
\ee
This temperature is inversely proportional to the mass of the black hole, 
and since the radiation reduces the mass, the radiation temperature will 
increase with time, as the black hole evaporates. For black holes of stellar 
size, however, one finds $T_{BH}~\lsim~2 \times 10^{-8}$ $^{\circ}$K, which
is many orders of magnitude below the 2.7 $^{\circ}$K cosmic microwave 
background, and hence not detectable. 

\medskip

It is instructive to consider the Schwarzschild radius of a typical hadron, 
assuming a mass $m \sim 1$ GeV:
\be
R_g^{had} \simeq 1.3 \times 10^{-38}~{\rm GeV}^{-1} \simeq 2.7 \times
10^{-39}~{\rm fm}.
\ee

To become a gravitational black hole, the mass of the hadron would thus
have to be compressed into a volume more than $10^{100}$ times smaller
than its actual volume, with a radius of about 1 fm. On the other hand,
if instead we increase the interaction strength from gravitation to strong
interaction \cite{C-R}, we gain in the resulting ``strong'' Schwarzschild 
radius $R_s^{had}$ a factor
\be
{\alpha_s \over Gm^2},
\ee
where $\alpha_s$ is the dimensionless strong coupling and $Gm^2$ the
corresponding dimensionless gravitational coupling for the case in question. 
This leads to
\be
R_s^{had} \simeq {2 \alpha_s \over m};  
\ee
with the effective value of $\alpha_s \sim {\mathcal O}(1)$ we thus 
get $R_s^{had} \sim {\mathcal O}(1)$ fm.\footnote{In fact, some studies 
\cite{alpha} indicate that at large distances, the strong coupling freezes 
at $\alpha_s \simeq 3$; in that case the corresponding radius becomes  
$R_s^{had} \simeq 1$ fm.} In other words, the confinement radius of a 
hadron is about the size of its ``strong'' Schwarzschild radius, so that 
we could consider quark confinement as the strong interaction version of 
the gravitational confinement in black holes \cite{C-R,Salam}. 

\medskip

We had seen that the mass of a black hole grows linearly with the 
event horizon, $M=(1/2G)R$, so that in gravitation $1/2G$ plays the role 
of the string tension in strong interaction physics. The replacement
$GM^2 \to \alpha_s$ here leads to
\be
\sigma \simeq {m^2 \over 2 \alpha_s} \simeq 0.16~{\rm GeV}^2,
\ee
if one uses the mentioned effective saturation value $\alpha_s\simeq 3$ 
\cite{alpha}. The value of $\alpha_s \sim 1$ thus gives a reasonable 
string tension as well as a reasonable
radius.  

\subsection{Quasi-Abelian case}

The appearence of an event horizon occurs in general relativity through the
modification of the underlying space-time structure by the gravitational
interaction. Such modifications have also been discussed for other
interactions. In particular, it was noted that in electrodynamics, non-linear 
in-medium effects can lead to photons propagating along geodesics which are 
not null in Minkowski space-time; this can even lead to photon trapping, 
restricting the motion of photons to a compact region of space \cite{Novello}. 
Thus, an effective Lagrangian ${\cal L}(F)$ depending on a one-parameter 
background field, $F=F_{\mu \nu}F^{\mu \nu}$, results in a modified metric 
\be
g_{\mu \nu} = \eta_{\mu \nu} {\cal L}' - 4 F_{\alpha \mu} F^{\alpha}_{\nu}
{\cal L}'',
\label{g}
\ee
where the primes indicate first and second derivatives with respect to $F$.
Hence
\be
g_{00} = {\cal L}' - 4F {\cal L}'' =0
\label{g-0}
\ee
defines the radius of the compact region of the theory, i.e., the
counterpart of a black hole \cite{Novello}.

\medskip

QCD is an inherently non-linear theory, with the physical vacuum playing
the role of a medium \cite{TD-vac}. The general structure of the effective 
Lagrangian in a background field $F$, compatible with gauge invariance, 
renormalization group results \cite{Gross:1973id} and trace anomaly, has 
the unique form 
\cite{L-QCD}
\be
{\cal L}_{QCD} = {1\over 4} F_{\mu \nu}F^{\mu \nu} {g^2(0) \over g^2(gF)}
={1\over 4} F_{\mu \nu}F^{\mu \nu} {\epsilon(gF)}.
\ee
Here $\epsilon(gF)$ is the dielectric ``constant'' of the system in 
the presence of the background field; the $F$-dependence of $\epsilon(F)$
effectively turns the QCD vacuum into a non-linear medium. On a one-loop 
pertubative level we have
\be
{\epsilon(gF)} \simeq 1 - \beta_0 \left({g^2 \over 4\pi}\right)
\ln {\Lambda^2 \over gF},
\ee
where $\beta_0=(11N_c - 2N_f)/48\pi^2$, with $N_c$ and $N_f$ specifying
the number of colors and flavors, respectively. Using this form in
the formalism of \cite{Novello} leads to $g_t$ changing sign (i.e.,
$g_t=0$) at 
\be
gF^* = \Lambda^2 \exp\{-4\pi/\beta_0 g^2\},
\ee
indicating a possible horizon at $r^* \sim 1/\sqrt{gF^*}$ \cite{Raufeisen}.
It is clear that this line of argument can at best provide some hints,
since we used the lowest order perturbative form of the beta-function,
even though at the horizon perturbation theory will presumably break
down. Nevertheless, we believe that it suggests the possibility of
an event horizon for QCD; the crucial feature is the asymptotic freedom 
of QCD \cite{Gross:1973id}, which leads to $\epsilon <1$ and allows $g_{00}$ 
to vanish even without the external medium effects required in QED.

\subsection{Non-Abelian case}

Indeed, a different and more solid suggestion that in QCD there is an
event horizon comes from studying the theory on a curved background.
For gluodynamics, such a program is
discussed in \cite{KLT-Pom}. Classical gluodynamics is a
scale-invariant
theory, but quantum fluctuations break this invariance, with the
trace of
the energy-momentum tensor introducing non-perturbative effects,
associated
with the vacuum energy density $\E$.
It was shown \cite{MSh} that low-energy theorems can be used to
determine
the form of the effective Yang-Mills Lagrangian in a curved but
conformally
flat metric
\be
g_{\mu \nu}(x) = \eta_{\mu \nu} e^{h(x)},
\ee
where the dilaton field $h(x)$ is coupled to the trace of the energy
momentum tensor, $\theta_\mu^\mu$. The resulting action has the form
\be
S = \int d^4x~ \left[ {4\over 3}{\E\over m_G^2} e^h (\partial_\mu
h)^2
- \frac{1}{4}
(F_{\mu \nu}^a)^2 + e^{2h} (\E - \frac{1}{4}
\theta_{\mu}^{\mu}|_{pert}).\right];
\label{Q-action}
\ee
here $\E$ is the absolute value of the
energy density of the vacuum and $m_G$ the dilaton mass;
the trace of the energy-momentum tensor has been separated into
perturbative
and non-perturbative contributions,
\be\label{traceen}
\theta_\mu^\mu = \theta_{\mu}^{\mu}|_{pert}+<\theta_\mu^\mu> =
\theta_{\mu}^{\mu}|_{pert} - 4\E.
\ee
The crucial point for our considerations is that the first term of 
eq.\ (\ref{Q-action}) can be written as
\be
{3\over 2} e^h (\partial_\mu h)^2 = R \sqrt{-g},
\ee
defining $R$ as the Ricci scalar of the theory. Hence
eq.\ (\ref{Q-action}) has the structure of an Einstein-Hilbert
Lagrangian of gluodynamics in the presence of an effective
gravitation, 
\be
S_G = \int d^4x~\left[{\sqrt{-g}\over 8\pi G} R - \frac{1}{4}
(F_{\mu \nu}^a)^2 + e^{2h} (\E - \frac{1}{4}
\theta_{\mu}^{\mu}|_{pert}).\right];
\label{G-action}
\ee
where  $G$ is now given by
\be
{1\over G} = {64 \pi \over 3} {\E\over m_G^2}.
\ee
The relation $1/2G \to \sigma$ between $G$ and the string tension
conjectured above then
leads to
\be
\sigma =  {32 \pi \over 3} {\E\over m_G^2}.
\label{s-1}
\ee
On the other hand, the string tension is just the energy density of
the
vacuum times the transverse string area,
\be
\sigma =  \E\pi r_T^2.
\label{s-2}
\ee
Combining relations (\ref{s-1}) and (\ref{s-2}), we have
\be
r_T=\sqrt{32\over 3} {1\over m_G} \simeq 0.4~{\rm fm},
\label{rtrans}
\ee
using $m_G\simeq 1.5$ GeV for the scalar glueball mass. Eq.\
(\ref{rtrans})
thus gives us the transverse extension or horizon of the string.

\medskip

From eqs.\ (\ref{s-1}) or (\ref{s-2}) we can obtain a further
consistency
check. Given the glueball mass and the string tension $\sigma\simeq
0.16$
GeV$^2$, we find for the vacuum energy density
\be
\E \simeq {3 \over 32 \pi} \sigma\ m_G^2 \simeq 0.013~{\rm GeV}^4
\simeq 1.7 ~{\rm GeV/fm}^3.
\ee
This is the value for pure gluodynamics; since the energy density is
related to the trace
of the energy--momentum tensor by the relation (\ref{traceen}), and
\be
\theta_\mu^\mu = {\beta(g) \over 2 g}\ (F_{\mu \nu}^a)^2 \simeq - b
{g^2 \over 32 \pi^2}\ (F_{\mu \nu}^a)^2,
\ee
with the coefficient $b = 11 N_c - 2 N_f$ of the $\beta$-function, we
can estimate that for three-flavor QCD
\be
\E_{\rm QCD} = {11N_c - 2N_f \over 11 N_c} \E =                     
 
{9 \over 11} \E \simeq 0.01 ~{\rm GeV}^4,
\ee
\vskip0.1cm
which is in perfect agreement with the original value of the gluon
condensate \cite{SVZ}
\be
\left< {\alpha_s \over \pi}  G^2 \right> \simeq 0.012
~{\rm GeV}^4;
\ee
\vskip0.1cm
note that $\E_{\rm QCD} = 27/32  \ \left< (\alpha_s /\pi)  G^2 \right>$).

\vskip0.5cm

\section{Hyperbolic Motion and Hawking-Unruh Radiation}

\begin{figure}[h]
\centerline{\psfig{file=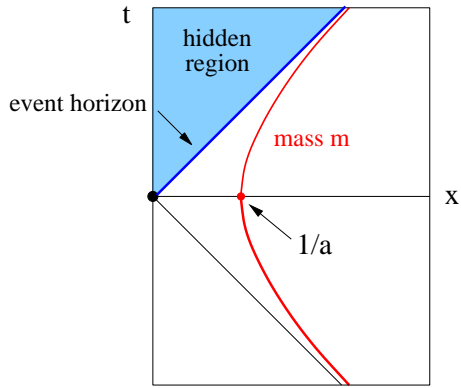,width=6cm}} 
\caption{Hyberbolic motion}
\label{W-L}
\end{figure}

In general relativity, the event horizon appeared as consequence of the
geometrized gravitational force, but its occurrence and its role for 
thermal radiation was soon generalized by Unruh \cite{Unruh}. 
A system undergoing uniform acceleration $a$ relative to a stationary 
observer eventually reaches a classical turning point and thus encounters 
an event horizon. Let us recall the resulting hyperbolic motion \cite{Pauli}. 
A point mass $m$ subject to a constant force $F$ satisfies the equation 
of motion
\be
{d \over dt}{mv \over \sqrt{1-v^2}} = F,
\label{e-motion} 
\ee
where $v(t)=dx/dt$ is the velocity, normalized to the speed of light $c=1$. 
This equation is solved by the parametric form through the so-called
Rindler coordinates,  
\be
x = \xi \cosh a\tau ~~~~~ t= \xi \sinh a \tau,
\label{rindler}
\ee
where $a=F/m$ denotes the acceleration in the instantaneous rest frame of 
$m$, and $\tau$ the proper time, with $d\tau=\sqrt{1-v^2} dt$. If we
impose the boundary condition that the velocity at $t=0$ vanishes,
we have $\xi=1/a$ and $x(t\!=\!0)=1/a$. The resulting
world line is shown in Fig.\ \ref{W-L}. It corresponds to the mass $m$
coming from $x=\infty$ at $t=-\infty$ at with a velocity arbitrarily close
to that of light, decelerating uniformly until it comes to rest at the
classsical turning point $x_H=-(1/a),t=0$. Subsequently, it accelerates 
again and
returns to $x=\infty$ at $t=\infty$, approaching the speed of light. 
For given $a$, the light cone originating at a distance $x_H=1/a$ away
from the turning point of $m$ defines a space-time region inaccessible to 
$m$: no photon in this region can (classically) ever reach $m$, in much 
the same way as photons cannot escape from a black hole. Here the 
acceleration is crucial, of course; if $m$ stops accelerating, it will
eventually become visible in the ``hidden region''.

\medskip

\medskip

The metric of such a accelerating system becomes in spherical coordinates
\cite{Gerlach} 
\be
ds^2 = \xi^2 a^2 d\tau^2 - d\xi^2 - \xi^2 \cosh^2 a \tau (d\theta^2
 + \sin^2 \theta d\phi^2),
\label{R-metric} 
\ee
which we want to compare to the black hole metric (\ref{metric}).
Making in the latter the coordinate transformation \cite{Terashima}
\be
\eta = {\sqrt g \over \kappa},
\ee
where the surface gravity $\kappa$ is given by
\be
\kappa = {1\over 2}~\!\left({\partial g \over \partial r}\right)_{r=R},  
\ee
we obtain for $r \to R$ the black hole form
\be
ds^2 =
 \eta^2 \kappa^2 dt^2 - d\eta^2 - R^2
(d\theta^2 + sin^2\theta d\phi^2).
\label{R-bh-metric} 
\ee
When we compare eqs.\ (\ref{R-metric}) and (\ref{R-bh-metric}),
it is evident that the system in uniform acceleration can be mapped
onto a spherical black hole, and vice versa, provided we identify
the surface gravity $\kappa$ with the acceleration $a$.

\medskip

\begin{figure}[h]
\centerline{\psfig{file=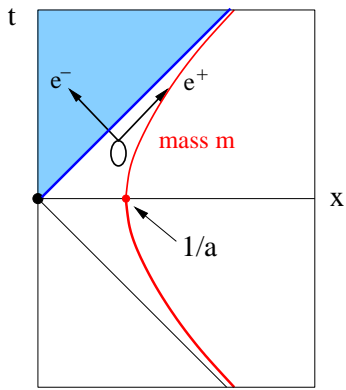,width=4.5cm}} 
\caption{Unruh radiation}
\label{U-P}
\end{figure}

The vacuum through which $m$ travels is, for a stationary observer, empty 
space. On a quantum level, however, it contains vacuum fluctuations. The 
accelerating mass $m$ can bring these on-shell, using up a (small) part of
its energy, so that for $m$ the vacuum 
becomes a thermal medium of temperature 
\be
T_U = {a \over 2 \pi}.
\label{T-U}
\ee
Consider such a fluctuation into an $e^+e^-$ pair, flying apart in
opposite directions. One electron is absorbed by the mass $m$, the
other penetrates into the ``hidden region'' and can never be detected by
$m$ (see Fig.\ \ref{U-P}. Since thus neither an observer on $m$ nor 
a stationary 
observer in the hidden region can ever obtain access to full information, 
each will register the observed radiation as thermal (Einstein-Podolsky-Rosen 
effect \cite{EPR}).
In other words: the accelerating mass $m$ sees the vacuum as a physical
medium of temperature $T_U$, while a stationary observer in the hidden
region observes thermal radiation of temperature $T_U$ as a consequence of
the passing of $m$. 

\medskip

\medskip

We here also mention that the entropy in the case of an accelerating 
system again becomes $1/4$ of the event horizon area, as in the black
hole case, so that also here the correspondence remains valid 
\cite{Laflamme}.

\medskip

In the case of gravity, we have the force
\be
F= m~\!a = G {m~\! M \over R^2},
\label{FG}
\ee
on a probe of mass $m$. With $R = 2~G~\!M$ for the (Schwarzschild)
black hole radius, we have $a=1/(4~G~\! M)$ for the acceleration at 
the event horizon and hence the Unruh temperature (\ref{T-U}) leads
back to the Hawking temperature of eq.\ (\ref{T-BH}). 

\medskip

In summary, we note that constant acceleration leads to an event horizon, 
which can be surpassed only by quantum tunnelling and at the expense of 
complete information loss, leading to thermal radiation as the only 
resulting signal.

\vskip0.5cm

\section{Pair Production and String Breaking}

In the previous section, we had considered a classical object, the mass $m$,
undergoing accelerated motion in the physical vacuum; because of quantum
fluctuations, this vacuum appears to $m$ as a thermal medium of temperature
$T_U$. In this section, we shall first address the modifications which arise 
if the object undergoing accelerated motion is itself a quantum system,
so that in the presence of a strong field it becomes unstable under pair 
production. Next we turn to the specific additional features which come 
in when the basic constituents are subject to color confinement and can 
only exist in color neutral bound states. 

\medskip 

\begin{figure}[h]
\centerline{\psfig{file=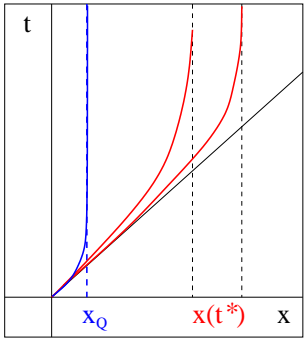,width=4cm}} 
\caption{Classical and quantum horizons in $\q$ separation}
\label{TP}
\end{figure}

As starting point, we consider two-jet $e^+e^-$ annihilation 
at cms energy $\sqrt s$, 
\be
e^+e^- \to \gamma^* \to \q \to~{\rm hadrons}.
\label{two-jet}
\ee  
The initially produced $\q$ pair flies apart, subject to the constant 
confining force given by the string tension $\sigma$; this  
results in hyperbolic motion \cite{Hosoya} of the type discussed in 
the previous section. At $t=0$, the $q$ and $\bar q$ separate with an 
initial velocity $v_0 = p/\sqrt{p^2 + m^2}$, where $p \simeq \sqrt s/2$ 
is the momentum of the primary constituents in the overall cms and $m$ 
the effective quark mass. We now have to solve Eq.\ (\ref{e-motion}) 
with this situation as boundary condition; the force
\be
F = \sigma,
\label{sigma}
\ee
is given by the string tension $\sigma$ binding the $\q$ system. The
solution is
\be
\tx = [1 - \sqrt{1-v_0\tt + \tt^2}]
\label{sol1}
\ee
with $\tx=x/x_0$ and $\tt=t/x_0$; here the scale factor 
\be
x_0 = {m \over \sigma} {1\over \sqrt{1-v_0^2}} = {1 \over a}~\gamma
\label{sol2}
\ee
is the inverse of the acceleration $a$ measured in the overall cms. The 
velocity becomes
\be
v(t) = {dx \over dt} = {(v_0/2) - \tt \over \sqrt{1 - v_0\tt + \tt^2}};
\label{sol3}
\ee
it vanishes for
\be
\tt^* = {v_0\over 2} ~\Rightarrow~ t^* = {v_0\over 2}~{m\over \sigma}~\gamma,
\label{sol4}
\ee
thus defining
\be
x(t^*) = {m\over \sigma}~\gamma~\left( 1 - \sqrt{1-(v_0^2/4)}\right) 
\simeq {\sqrt s \over 2 \sigma}
\label{sol5}
\ee
as classical turning point and hence as the classical event horizon measured 
in the overall cms (see Fig.\ \ref{TP}). 

Eq.\ (\ref{sol5}) allows the $q$ and the $\bar q$ to separate arbitrarily
far, provided the pair has enough initial energy; this clearly violates 
color confinement. Our mistake was to consider the $\q$ system as classical;
in quantum field theory, it is not possible to increase the potential 
energy of a given $\q$ state beyond the threshold value necessary to bring 
a virtual $\q$ pair on-shell. In QED, the corresponding phenomenon was 
addressed by Schwinger \cite{Schwinger}, who showed that in the presence 
of a constant electric field of strength $\cal E$ the probability of 
producing an electron-positron pair is given by
\be
P(M,{\cal E}) \sim \exp\{-\pi m_e^2 / e {\cal E}\},
\label{schwinger}
\ee
with $m_e$ denoting the electron mass and $e$ denoting the electric charge. 
This result is in fact a specific case of the Hawking-Unruh phenomenon,
as shown in \cite{K-T}.
In QCD, we expect a similar effect when the string tension exceeds the pair 
production limit, i.e., when
\be
\sigma~x > 2~\!m
\label{pair}
\ee
where $m$ specifies the effective quark mass. Beyond this point, any further 
stretching of the string is expected to produce a $\q$ pair with the 
probability
\be
P(M,\sigma) \sim \exp\{-\pi m^2 / \sigma\},
\label{string-pair}
\ee
with the string tension $\sigma$ replacing the electric field strength
$e\cal E$. This string breaking acts like a quantum event horizon 
$x_q = 2~\!m / \sigma$, which becomes operative long before the classical 
turning point is ever reached (see Fig.\ \ref{TP}). Moreover, the resulting 
allowed separation distance for our $\q$ pair, the color confinement 
radius $x_Q$, does not depend on the initial energy of the primary quarks. 

\medskip

There are some important differences between QCD and QED. In case of the 
latter, the initial electric charges which lead to the field $\cal E$ can 
exist independently in the physical vacuum, and the produced pair can be 
simply ionized into an $e^+$ and an $e^-$. In contrast, neither the primary
quark nor the constituents of the $\q$ pair have an independent existence, 
so that in string breaking color neutrality must be preserved. As a result,
the Hawking radiation in QCD must consist of $\q$ pairs, and these can
be produced in an infinite number of different excitation states of 
increasing mass and degeneracy. Moreover, the $\q$ pair spectrum is
itself determined by the strength $\sigma$ of the field, in contrast
to the exponent $m_e^2/\cal E$ in eq.\ (\ref{schwinger}), where the value
of $\cal E$ has no relation to the electron mass $m_e$.  

\medskip
 
\begin{figure}[h]
\centerline{\psfig{file=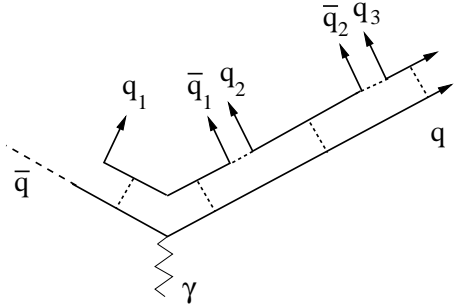,width=6cm}} 
\caption{String breaking through $\q$ pair production}
\label{breaking}
\end{figure}

Hadron production in $e^+e^-$ annihilation is believed to proceed in 
the form of a self-similar cascade \cite{bj,nus}. Initially, we have the
separating primary $\q$ pair, 
\be
\gamma \to [\q]
\label{bj1}
\ee
where the square brackets indicate color neutrality. When the energy of
the resulting color flux tube becomes large enough, a further pair 
$q_1\bar q_1$ is excited from the vacuum by two-gluon exchange (see
Fig.\ \ref{breaking}), 
\be
\gamma \to [q [\bar q_1 q_1] \bar q].
\label{bj2}
\ee
Although the new pair is at rest in the overall cms, each of its 
constituents has a transverse momentum $k_T$ determined, through 
the uncertainty relation, by the transverse dimension $r_T$ of the flux 
tube. The slow $\bar q_1$ now screens the fast primary $q$ from its 
original partner $\bar q$, with an analoguous effect for the $q_1$ and
the primary antiquark. To estimate the $\q$ separation distance at the
point of pair production, we recall that the thickness of the 
flux tube connecting the $\q$ pair is in string theory given by 
\cite{Luescher}
\be
r_T^2 = {2\over \pi \sigma} \sum_{k=0}^K {1 \over 2k+1},
\label{thick}
\ee
where $K$ is the string length in units of an intrinsic vibration measure.
Lattice studies \cite{Bali} show that for strings in the range of 1 - 2 fm, 
the first string excitation dominates, so that we have
\be
r_T = c_0 \sqrt{2\over \pi \sigma},
\label{thick-0}
\ee
with $c_0 \simeq 1$ or slightly larger. Higher excitations lead to a 
greater thickness and eventually to a divergence (the ``roughening'' 
transition). From the uncertainty relation we then have
\be
k_T = {1 \over c_0}\sqrt{\pi \sigma \over 2}.
\label{kt}
\ee 
With this transverse energy is included in eq.\ (\ref{pair}), we obtain 
for the pair production separation $x_Q$ 
\be
\sigma x_q = 2\sqrt{m^2 + k_T^2} ~~\Rightarrow~~x_q \simeq 
{2~\over\sigma}\sqrt{m^2 + (\pi \sigma/2~\! c_0^2)} \simeq 
\sqrt{2\pi \over \sigma c_0^2}
\simeq 1~{\rm fm},
\label{x-H}   
\ee
with $\sigma = 0.2$ GeV$^2$, $m^2 \ll \sigma$, and $c_0 \simeq 1$. 

\medskip

\medskip

Once the new pair is present, we have a color-neutral system 
$q\bar q_1 q_1 \bar q$; but since there is a sequence of connecting
string potentials $q \bar q_1$, $\bar q_1 q_1$ and $q_1 \bar q$,
the primary string is not yet broken. To achieve that, the binding
of the new pair has to be overcome, i.e., the $q_1$ has to tunnel through
the barrier of the confining potential provided by $\bar q_1$, and 
vice versa. Now the $q$ excerts a longitudinal force on the $\bar q_1$, 
the $\bar q$ on the $q_1$, resulting in a longitudinal acceleration and 
ordering of $q_1$ and $\bar q_1$. When (see Fig.\ \ref{breaking})
\be
\sigma x(q_1\bar q_1) = 2\sqrt{m^2 + k_T^2},
\ee
the $\bar q_1$ reaches its $q_1\bar q_1$ horizon; on the other hand, when
\be
\sigma x(q \bar q_1) = 2\sqrt{m^2 + k_T^2},
\ee
the new flux tube $q \bar q_1$ reaches the energy needed to produce a
further pair $q_2 \bar q_2$. The $\bar q_2$ screens the primary $q$
from the $q_1$ and forms a new flux tube $q \bar q_2$. At this point,
the original string is broken, and the remaining pair $\bar q_1 q_2$ 
form a color neutral bound state which is emitted as Hawking radiation
in the form of hadrons, with the relative weights of the different states
governed by the corresponding Unruh temperature. The resulting pattern is 
schematically illustrated in Fig.\ \ref{breaking}.

\medskip

To determine the temperature of the hadronic Hawking radiation, we 
return to the original pair excitation process. To produce a quark of
momentum $k_T$, we have to bring it on-shell and change its velocity
from zero to $v = k_T / (m^2 + k_T^2)^{1/2} \simeq 1$. This has to
be achieved in the time of the fluctuation determined by the virtuality
of the pair, $\Delta \tau = 1/\Delta E \simeq 1/ 2k_T$. The resulting
acceleration thus becomes
\be
a = {\Delta v \over \Delta \tau} \simeq 2~\!k_T \simeq 
\sqrt{2 \pi \sigma}/c_0 \simeq 1~{\rm GeV},
\ee
which leads to 
\be
T_Q = {a\over 2 \pi} \simeq {1\over c_0} \sqrt{\sigma \over 2 \pi} 
\simeq 160 - 180~{\rm MeV}
\label{T-H}
\ee
for the hadronic Unruh temperature. It governs the momentum distribution
and the relative species abundances of the emitted hadrons.

\medskip

\begin{figure}[h]
\centerline{\psfig{file=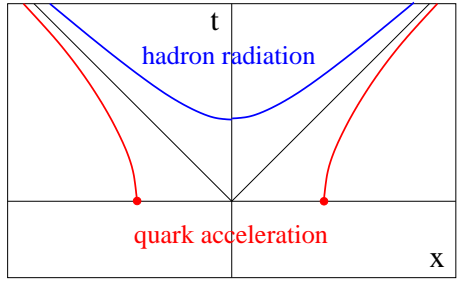,width=6cm}} 
\caption{Quark acceleration and hadronization world lines}
\label{world-a}
\end{figure}

A given step in the evolution of the hadronization cascade of a primary
quark or antiquark produced in $e^+e^-$ annihilation thus involves several
distinct phenomena. The color field created by the separating $q$ and 
$\bar q$ produces a further pair $q_1\bar q_1$ and then provides an
acceleration of the $q_1$, increasing its longitudinal momentum. When
it reaches the $q_1\bar q_1$ confinement horizon, still another pair
$q_2\bar q_2$ is excited; the state $\bar q_1 q_2$ is emitted as a hadron,
the $\bar q_2$ forms together with the primary $q$ a new flux tube. This
pattern thus step by step increases the longitudinal momentum of the
``accompanying'' $\bar q_i$ as well as of the emitted hadron. This, 
together with the energy of the produced pairs, causes a corresponding
deceleration of the primary quarks $q$ and $\bar q$, in order to maintain
overall energy conservation. In Fig.\ \ref{world-a}, we show the world
lines given by the acceleration $\bar q_i \to \bar q_{i+1}$ 
($q_i \to q_{i+1}$) and that of formation threshold of the hadrons 
$\bar q_i q_{i+1}$ and the corresponding antiparticles. 

\medskip

\begin{figure}[h]
\centerline{\psfig{file=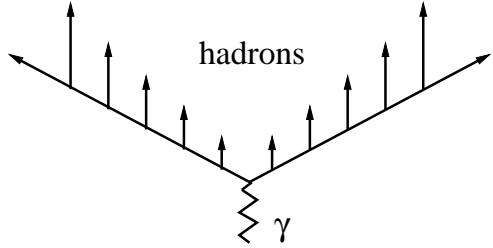,width=6.5cm}} 
\caption{Hadronization in $e^+e^-$ annihilation}
\label{multi}
\end{figure}

The energy loss and deceleration of the primary quark $q$ in this self-similar 
cascade, together with the acceleration of the accompanying partner $\bar q_i$
from the successive pairs brings $q$ and $\bar q_i$ closer and closer to 
each other in momentum, from an initial separation $q \bar q_1$ of 
$\sqrt s/2$, until they finally are combined into a hadron and the 
cascade is ended. The resulting pattern is shown in Fig.\ \ref{multi}.

\medskip

The number of emitted hadrons, the multiplicity $\nu(s)$, follows quite
naturally from the picture presented here. The classical string length,
in the absence of quantum pair formation, is given by the classical
turning point determined in eq.\ (\ref{sol5}). The thickness of a flux
tube of such an ``overstretched'' string is known \cite{Luescher}; from
eq.\ (\ref{thick}) we get
\be
R_T^2 = {2 \over \pi \sigma} \sum_{k=0}^K {1 \over 2k + 1}
\simeq {2 \over \pi \sigma} \ln 2K,
\ee
where $K$ is the string length. From eq.\ (\ref{sol5}) we thus get
\be
R_T^2 \simeq {2 \over \pi \sigma} \ln \sqrt s
\ee
for the flux tube thickness in the case of the classical string length. In 
parton language, the logarithmic growth of the transverse hadron size is due 
to parton random walk ("Gribov diffusion" \cite{Gribov}); this phenomenon 
is responsible for diffraction cone shrinkage in high--energy hadron 
scattering. 

\medskip

Because of pair production, the string breaks whenever it is stretched 
to the length $x_q$ given in eq.\ (\ref{x-H}); its thickness $r_T$ at this 
point is given by eq.\ (\ref{thick}). The multiplicity can thus be 
estimated by the ratio of the corresponding classical to quantum transverse 
flux tube areas,
\be
\nu(s) \sim {R_T^2 \over r_T^2} \sim \ln \sqrt s,
\ee 
and is found to grow logarithmically with the $e^+e^-$ annihilation energy,
as is observed experimentally over a considerable range. 

\medskip

We note here that in our argumentation we have neglected parton evolution,
which would cause the emitted radiation (e.g., $\bar q_1 q_2$ in Fig.\
\ref{breaking}) to start another cascade of the same type. Such 
evolution effects result eventually in a stronger increase of the 
multiplicity. The formation of a white hole does not affect the production
of hard processes at early times (e.g., multiple jet production), which
is responsible for an additional growth of the measured multiplicity. 

\medskip

A further effect we have not taken into account here is parton saturation.
At sufficiently high energy, stronger color fields can lead to gluon
saturation and thus to a higher temperature determined by the saturation
momentum \cite{K-T}. The resulting system then expands and hadronizes at 
the universal temperature determined by the string tension.

\medskip

It interesting to compare the separation of two energetic light quarks,
as we have considered here, with that of two static heavy quarks $Q$ and 
$\bar Q$. From quarkonium studies it is known that
\be
2(M_D-m_c) \simeq 2(M_B -m_b) \simeq 1.2~{\rm GeV},
\ee 
where $M_D(M_B)$ and $m_c(m_b)$ are the masses of open charm (beauty)
mesons and of the corresponding charm (beauty) quarks, respectively.
The energy needed to separate a heavy $\Q$ pair thus is independent of
the mass of the heavy quarks, indicating that the string breaking 
involved here is really a consequence of the vacuum, through $\q$
pair excitation. With
\be
\sigma x_Q \simeq 1.2~{\rm GeV} ~\Rightarrow~x_Q \simeq 1.2~{\rm fm}
\ee
we find that the resulting separation threshold for pair excitation agrees
well with that found above in eq.\ (\ref{x-H}). Lattice QCD studies lead 
to similar results. 

\medskip

\begin{figure}[h]
\centerline{\psfig{file=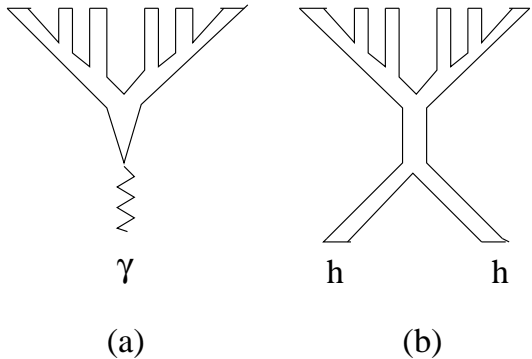,width=7cm}} 
\caption{``White hole'' structure in $e^+e^-$ annihilation (a) and 
hadronic collisions (b)}
\label{wh}
\end{figure}

\begin{figure}[h]
\centerline{\psfig{file=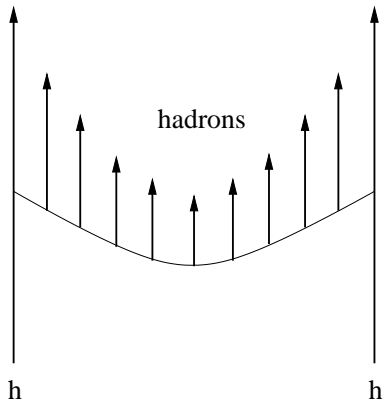,width=5cm}} 
\caption{Hadronization in hadron-hadron
collisions}
\label{hadromulti}
\end{figure}

Up to now, we have considered hadron production in $e^+e^-$ annihilation,
in which the virtual photon produces a confined colored $\q$ pair as a
``white hole''. Turning now to hadron-hadron collisions, we note that here
two incident white holes combine to form a new system of the same kind,
as schematically illustrated in Fig.\ \ref{wh}. Again the resulting string 
or strong color field produces a sequence of $\q$ pairs of increasing cms 
momentum, leading to the well-known multiperipheral hadroproduction 
cascade shown in Fig.\ \ref{hadromulti}. We recall here the comments
made above concerning parton evolution and saturation; in hadronic
collisions as well, these phenomena
will affect the multiplicity, but not the relative abundances.

\medskip

In the case of heavy ion collisions, two new elements enter. The resulting 
systems could now have an overall baryon number, up to $B=400$ or more. To
take that into account, we need to consider the counterpart of charged
black holes. Furthermore, in heavy ion collisions the resulting hadron 
production can be studied as function of centrality, and peripheral
collisions could lead to an interaction region with an effective
overall angular momentum. Hence we will also consider rotating black 
holes. In the next section, we then summarize the relevant features 
of black holes with $Q\not= 0,~J\not=0$.

\vskip0.5cm

\section{Charged and Rotating Black Holes}

As mentioned, an outside observer the only characteristics of a black
hole are its mass $M$, its electric charge $Q$, and its spin or angular
momentum $J$. Hence any further observables, such as the event horizon
or the Hawking temperature, must be expressable in terms of these three
quantities.

\medskip

\begin{figure}[h]
\centerline{\psfig{file=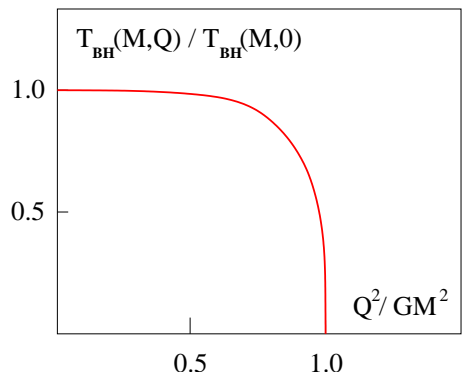,width=6cm}} 
\caption{Radiation temperature for a charged black hole}
\label{R-N}
\end{figure}

The event horizon of a black hole is created by the strong gravitational
attraction, which leads to a diverging Schwarzschild metric at a certain 
value of the spatial extension $R$. Specifically, the invariant space-time 
length element $ds^2$ is at the equator given by
\be
ds^2 = ( 1 - 2GM/R)~\!dt^2 - {1\over 1- 2GM/R}~\!dr^2,
\label{Schwarz}
\ee
with $r$ and $t$ for flat space and time coordinates; it is seen to diverge at
the Schwarzschild radius $ R_S = 2GM$. If the black hole has a net electric 
charge $Q$, the resulting Coulomb repulsion will oppose and hence weaken 
the gravitational attraction; this will in turn modify the event horizon.
As a result, the corresponding form (denoted as Reissner-Nordstr\"om metric) 
becomes
\be
ds^2 = (1 - 2GM/R + GQ^2/R^2)~\!dt^2 - {1\over 1- 2GM/R + GQ^2
/R^2}~\!dr^2.
\label{RN}
\ee
For this, the divergence leads to the smaller Reissner-Nordstr\"om radius
\be
R_{RN} = GM~\!(1 + \sqrt{1 - Q^2/GM^2}),
\label{RNradius}
\ee
which reduces to the Schwarzschild radius $R_S$ for $Q=0$. 
The temperature of the Hawking radiation now becomes \cite{Ruf,I-U-W}   
\be
T_{BH}(M,Q) = T_{BH}(M,0)~\!\left\{{4~\sqrt{1 - Q^2/ GM^2} 
\over (1 + \sqrt{1- Q^2/GM^2}~\!)^{~\!2}}\right\}; 
\label{T-Q}
\ee
its functional form is illustrated in Fig.\ \ref{R-N}. We note that with
increasing charge, the Coulomb repulsion weakens the gravitational field
at the event horizon and hence decreases the temperature of the corresponding
quantum excitations. As $Q^2 \to GM^2$, the gravitational force is fully
compensated and there is no more Hawking radiation.

\medskip

\begin{figure}[h]
\centerline{\psfig{file=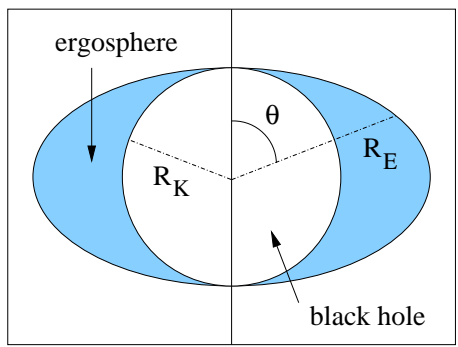,width=6cm}} 
\caption{Geometry of a rotating black hole}
\label{ergo}
\end{figure}

In a similar way, the effect of the angular momentum of a rotating black
hole can be incorporated. It is now the centripetal force which counteracts
the gravitational attraction and hence reduces its strength. The resulting  
Kerr metric must take into account that in this case the rotational symmetry 
is reduced to an axial symmetry, and with $\theta$ denoting the angle relative 
to the polar axis $\theta=0$, it is (at fixed longitude) given by
\be
ds^2 = \left( 1 - {2GMR \over R^2 + j^2 \cos^2 \theta}\right) dt^2 
- {R^2 + j^2 \cos^2 \theta \over R^2 - 2GMR + j^2}~\!dr^2 
- (R^2 + j^2 \cos^2 \theta)~\!d\theta^2.
\label{Kerr}
\ee
The angular momentum of the black hole is here specified by the parameter
$j=J/M$; for $a=0$, we again recover the Schwarzschild case. The general
situation is now somewhat more complex, since eq.\ (\ref{Kerr})
leads to two different divergence points. The solution
\be
R_K = GM~\!(1 + \sqrt{1 - j^2/(GM)^2}~\!)
\label{RK}
\ee 
defines the actual event horizon, corresponding to absolute confinement.
But the resulting black hole is now embedded in a larger ellipsoid
\be
R_E = GM~\!(1 + \sqrt{1 - [j^2/(GM)^2]\cos^2 \theta}),
\label{RE}
\ee 
as illustrated in Fig.\ \ref{ergo}. The two surfaces touch at the poles, and 
the region between them is denoted as the ergosphere. Unlike the black hole
proper, communication between the ergosphere and the outside world is
possible. Any object in the ergosphere will, however, suffer from the
rotational drag of the rotating black hole and thereby gain momentum. 
We shall return to this shortly; first, however, we note that the 
temperature of the Hawking radiation from a rotating black hole becomes
\be
T_{BH}(M,J) = T_{BH}(M,0)~\!\left\{{2 \sqrt{1 - j^2/(GM)^2} 
\over 1 + \sqrt{1- j^2/(GM)^2} }\right\}. 
\label{T-J}
\ee
For a non-rotating black hole, with $j=0$, this also reduces to the 
Hawking temperature for the Schwarzschild case.  

\medskip

To illustrate the effect of the ergosphere, imagine radiation from a  
Schwarzschild black hole emitted radially outward from the event horizon. 
In the case of a Kerr black hole, such an emission is possible only along 
the polar axis; for all other values of $\theta$, the momentum of the 
emitted radiation (even light) will increase due to the rotational drag 
in the ergosphere. This effect ceases only once the radiation leaves
the ergosphere. Since the amount of drag depends on $\theta$, the momentum 
of the radiation emitted from a rotating black hole, as measured at large 
distances, will depend on the latitude at which it is emitted and increase 
from pole to equator. 

\medskip

Finally, for completeness, we note that for black holes with both spin and 
charge (denoted as Kerr-Newman), the event horizon is given by
\be
R_{KN} = GM~\!(1 + \sqrt{1 - [Q^2/GM^2] - [j^2/(GM)^2]}),
\label{KN}
\ee 
and the radiation temperature becomes
\cite{Ruf,I-U-W}   
\be
T_{BH}(M,Q,J) = T_{BH}(M,0,0)~\left\{{4 \sqrt{1 - (GQ^2 + j^2)/(GM)^2} 
\over (1 + \sqrt{1- (GQ^2 +j^2)/(GM)^2}~\!)^{~\!2} + j
^2/(GM)^2}\right\}. 
\label{T-QJ}
\ee
The decrease of $T_{BH}$ for $Q\not= 0, J\not= 0$ expresses the fact that 
both the Coulomb repulsion and the rotational force counteract the 
gravitational attraction, and if they win, the black hole is dissolved. 

\medskip

The dependence of a black hole on its basic properties $M,Q,J$ is very
similar to the dependence of a thermodynamic system on a set of 
thermodynamic observables. The first law of thermodynamics can be
written as
\be\label{thermodyn}
dE = TdS + \phi dQ + \omega dJ,
\ee
expressing the variation of the energy with entropy $S$, charge $Q$ and 
spin $J$; here $\phi$ denotes the electrostatic potential per charge and
$\omega$ the rotational velocity. The corresponding relation in black hole 
thermodynamics becomes
\be\label{bhthermodyn}
dM = T_{BH} dS_{BH} + \Phi dQ + \Omega dJ,
\ee
where the entropy $S_{BH}$ is defined as the area of the event horizon,
\be
S_{BH} = {\pi (R_{KN}^2 + j^2)\over G}.
\ee
The temperature is given by eq.\ (\ref{T-QJ}), and
\be
\Phi = {4 \pi Q R_{RN} \over G~\!S_{BH}}, ~~~\Omega = {4\pi a \over S_{BH}}  
\label{pot}
\ee
specify the electrostatic potential $\Phi$ and the rotational velocity
$\Omega$. 

\medskip

The considerations of this section were for spherical black holes.
As seen above, such objects are in fact equivalent to uniformly
accelerating systems. An application to actual high energy collisions
involves a further assumption. Thermal Hawking-Unruh radiation arises
already from a single $\Q$ system, as seen above in the discussion of
$e^+e^-$ annihilation. If we treat the systems produced in heavy ion
collisions as black holes of an overall baryon number and/or an overall
spin, we are assuming that the collision leads to a large-scale collective
system, in which each accelerating parton is affected by totality of
the other accelerating partons. This assumption clearly goes beyond
our event horizon conjecture and, in particular, it need not be correct
in order to obtain thermal hadron production.   

\vskip0.5cm

\section{Baryon Density and Angular Momentum}

\subsection{Vacuum Pressure and Baryon Repulsion}

We now want to consider the extension of charged black hole physics to color 
confinement in the case of collective systems with a net baryon number.
In eq.\ (\ref{T-Q}) we had seen that the reduction of the gravitational
attraction by Coulomb repulsion in a charged black hole modifies the
event horizon and hence in turn also the temperature of Hawking radiation. 
The crucial quantity here is the ratio $Q^2/GM^2$ of the repulsive
overall Coulomb force, $Q^2/R^2$, to the attractive overall gravitational 
force, $GM^2/R^2$, at the horizon.

\medskip

In QCD, we have a ``white'' hole containing colored quarks, confined 
by chromodynamic forces or, equivalently, by the pressure of the physical 
vacuum. If the system has a non-vanishing overall baryon number, the 
baryon-number dependent interaction will also affect the forces at the 
event horizon. The simplest instance of such a force the repulsion between 
quarks due to Fermi statistics, but more generally, there will be repulsive
effects of the type present in cold dense baryonic matter, such as neutron
stars. The resulting 
pressure will modify the confinement horizon and hence lead to a 
corresponding modification of the Hawking-Unruh temperature of 
hadronization. 

\medskip

By using the conjectured correspondence between black hole thermodynamics 
and the thermodynamics of confined color charges, we translate black hole 
mass, charge and gravitational constant into white hole energy, net baryon 
number and string tension, 
\be
\{ M, Q, G \} \leftrightarrow \{E, B, 1/ 2 \sigma \}.
\ee
Hence eq.\ (\ref{T-Q}) leads us to the relation
\be
T_Q(B) = T_Q(B=0)~\!\left\{{4~\sqrt{1 - 2 \sigma~\! B^2 / E^2} 
\over (1 + \sqrt{1- 2 \sigma~\! B^2 /E^2}~\!)^{~\!2}}\right\}; 
\label{T-B}
\ee
for the dependence of the hadronization temperature on the ratio of
net baryon number $B$ and energy $E$, with $T_Q(B=0)$ given by 
eq.\ (\ref{T-H}). Its functional form is the same as that illustrated 
in Fig.\ \ref{R-N}.

\medskip

It would be interesting to test the prediction (\ref{T-B}) against  
experimental data; one could identify $B$ with the net baryon number 
per unit rapidity $dN_B/dy$ and $E$ with the total transverse 
energy per unit rapidity $dE_T/dy$. The reduction of the hadronization
temperature with baryon number could thus occur in two ways. A sufficient
decrease of the collision energy, e.g.\ from peak SPS to AGS energy,
will strongly reduce $dE_T/dy$, while $dN_B/dy$ is not affected as much.
This leads to the known decrease of $T(\mu_B)$ with increasing 
$\mu_B$ \cite{BCKSR},
and it will be interesting to see if the form (\ref{T-B}) agrees with
the observed behaviour. A second, novel possibility would be to consider 
hadrochemistry as a function of rapidity. At peak SPS energy, $dN_B/dy$
remains essentially constant out to about $y=2$, while $dE_T/dy$ drops
by more than a factor of two from $y=0$ to $y=2$ \cite{Sikler}. A similar 
behaviour
occurs at still lower collision energies. Hence it would seem worthwhile
to check if an abundance analysis at large $y$ indeed shows the expected
decrease of the hadronization temperature. 

\subsection{Angular Momentum and Non-Central Collisions}

\begin{figure}[h]
\centerline{\psfig{file=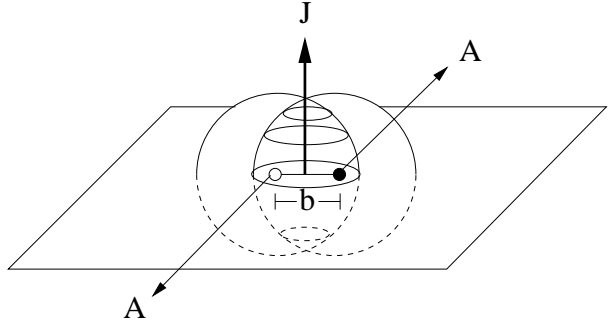,width=8cm}} 
\caption{Rotating interaction region in non-central $AA$ collision} 
\label{spin}
\end{figure}

The dependence of Hawking radiation on the angular momentum of the emitting
system introduces another interesting aspect for the ``white hole 
evaporation'' we have been considering.  Consider a nucleus-nucleus collision
at non-zero impact parameter $b$. If the interaction is of collective 
nature, the resulting interaction system may have some angular momentum 
orthogonal to the reaction plane (see Fig.\ \ref{spin}). In central 
collisions, this will not be the case, nor for extremely peripheral ones,
where one expects essentially just individual nucleon-nucleon collisions 
without any collective effects. 

\medskip

If it possible to consider a kinematic region in which the interacting
system does have an overall spin, then the resulting Hawking radiation
temperature should be correspondingly reduced, as seen in eq.\ (\ref{T-J}).
The effect is not so easily quantified, but simply a reduction of the
hadronization temperature for non-central collisions would quite indicative.
Such a reduction could appear only in the temperature determined by the
relative abundances, since, as we shall see shortly, the transverse 
momentum spectra should show modifications due to the role of the 
ergosphere. 

\medskip

\begin{figure}[h]
\vskip-5.4cm
\hfill{\psfig{file=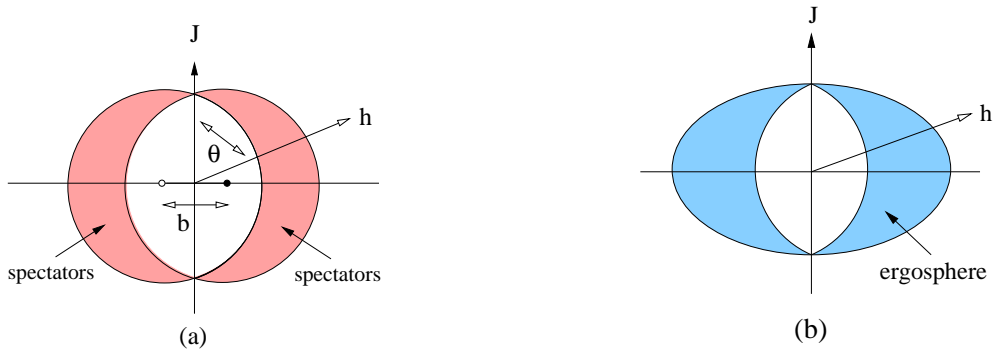,width=4.7cm}}~~~~~~~~~~
\bigskip
\caption{Transverse plane view of a non-central $AA$ collision} 
\label{ellip}
\end{figure}

We next turn to the momentum spectrum of the Hawking radiation emitted
from a rotating white hole. As discussed in section 4, such radiation
will exhibit an azimuthal asymmetry due to the presence of the ergosphere,
which by its rotation will affect the momentum spectrum of any passing
object. At the event horizon, the momentum of all radiation is determined 
by the corresponding Hawking temperature (\ref{T-J}); but the passage of 
the ergosphere adds rotational motion to the emerging radiation and
hence increases its momentum. As a result, only radiation emitted 
directly along the polar axis will have momenta as specified by the
Hawking temperature; with increasing latitude $\theta$ (see Fig.\ 
\ref{ellip})a, the rotation will increase the radiation momentum up to 
a maximum value in the equatorial plane. 

\medskip

\begin{figure}[h]
~~~~~~~~~~{\psfig{file=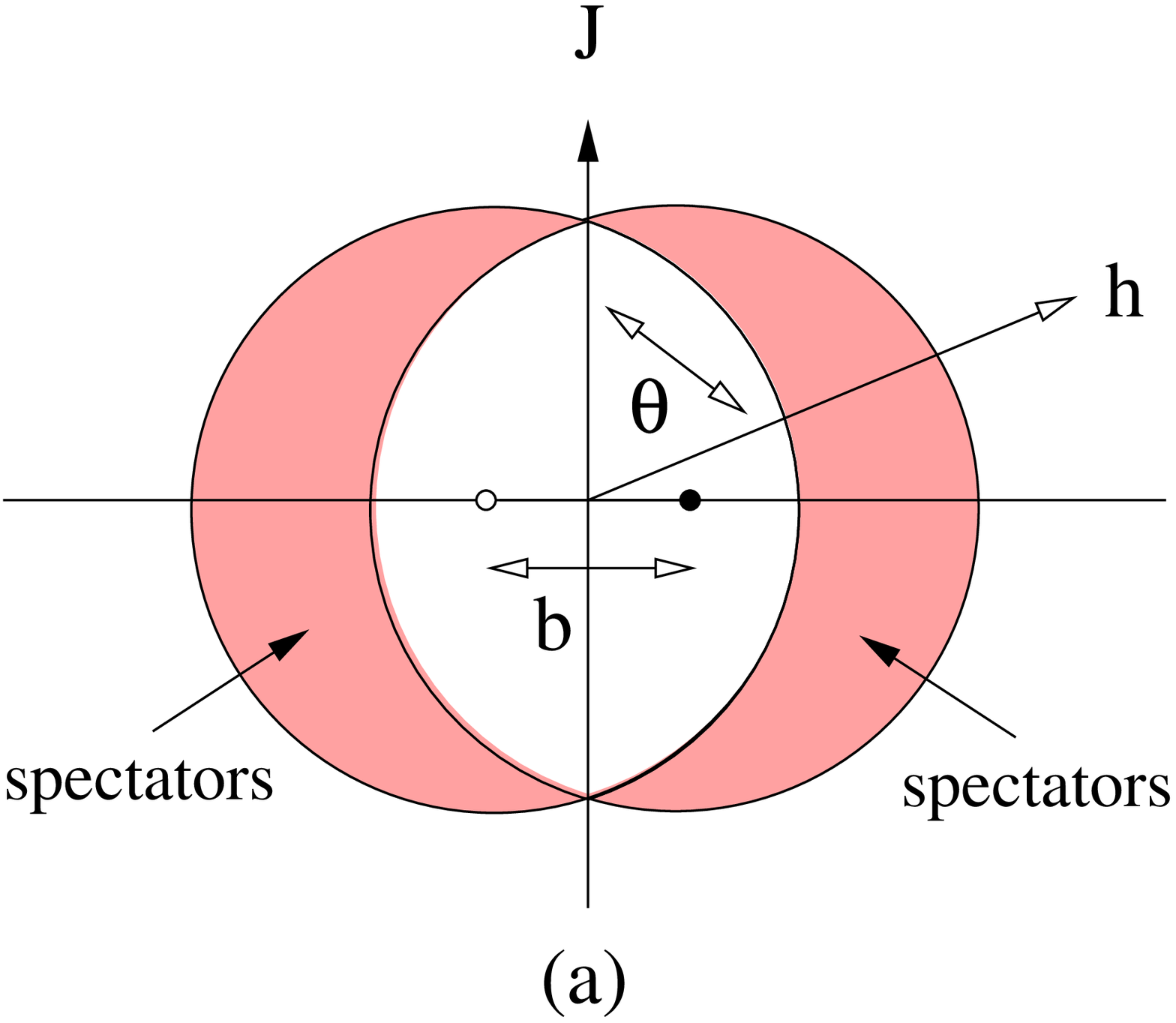,width=5cm}} 
\end{figure}

\medskip

Hawking radiation from a rotating source thus leads for nuclear collisions 
quite naturally to what in hydrodynamic studies is denoted as elliptic flow. 
It is interesting to note that both scenarios involve collective effects:
while in hydrodynamics, it is assumed that non-central collisions lead to
an azimuthally anisotropic pressure gradient, we have here assumed that
such collisions lead to an overall angular momentum of the emitting 
system.

\medskip

Concluding this section we emphasize that the results obtained here
for the Hawking temperatures of systems with finite baryon density 
or with an effective overall spin depend crucially on the assumption
of collectivity. If the different nucleon-nucleon interactions in
a heavy ion collision do not result in sufficiently collective
behavior, the corresponding modifications of $T_q$ do not apply.
In the case of black holes with spin, we moreover have no way to
 relate in a quantitative way centrality and overall spin. 
Both cases do show, however, that such extensions lead to qualitatively 
reasonable modifications.

\vskip0.5cm

\section{Temperature and Acceleration Limits}

We had seen that the underlying confinement dynamics of high energy hadron
collisions and $e^+e^-$ annihilation led to a limit on the acceleration
(or the corresponding deceleration) in the self-similar hadronization
cascade - a limit which can be specified in terms of the string tension.
In turn, this led to a limiting Unruh hadronization temperature
\be
T_Q \simeq \sqrt{\sigma \over 2 \pi}.
\label{Tunruh}
\ee
We emphasize that a Hawking-Unruh temperature as such can {\sl a priori}
have any value; it is the universal limit on the acceleration that leads
to a universal temperature for the emitted hadron radiation.

\medskip

In the study of strongly interacting matter, temperature limits are 
well-known and arise for an ideal gas of different composite constituents 
(``resonances'' or ``fireballs'' of varying mass $M$), if the composition 
law provides a sufficiently fast increase of the degeneracy $\rho(M)$ with $M$.
If the number of states of a constituent of mass $M$ grows exponentially,
\be  
\rho(M) \sim M^{-a} \exp\{bM\},
\label{expo}
\ee
with constants $a$ and $b$, then the grand canonical partition function
for an ideal gas in a volume $V$
\be
{\cal Z}(T,V) = \sum_N {1 \over N!} \left[ {V \over (2\pi)^3} \int dM
\rho(M) \int d^3p~\exp\{-\sqrt{p^2+M^2}~/T\} \right]^N, 
\ee
diverges for
\be
T > T_H \equiv 1/b,
\ee
so that the Hagedorn temperature $T_H$ \cite{Hagedorn} constitutes an upper 
limit for the temperature of hadronic matter. 

\medskip

In the dual resonance model \cite{drm}, the resonance composition pattern
is governed by linearly rising Regge trajectories,
\be
\alpha' M^2_n = n + \alpha_0, ~~n=1,2,...,
\ee
in terms of the universal Regge slope $\alpha'\simeq 1$ GeV$^{-2}$ and
a constant (of order unity) specifying the family ($\pi, \rho, ...$).
For an ideal resonance gas in $D-1$ space and one time dimension, one then
obtains \cite{h-w}
\be
a={1\over 2} (D+1), ~~~b= 2\pi \sqrt{D\alpha' \over 6},
\ee
leading to the temperature limit
\be
T_R = {1\over 2 \pi} \sqrt{6 \over \alpha'D}.
\ee
In string theory, the Regge resonance pattern is replaced by string
excitation modes, retaining the same underlying partition structure,
with $\alpha'=1/2\pi \sigma$ relating Regge slope and string tension.
Hence we get
\be
T_R = \sqrt{6 \sigma \over 2 \pi D}
\ee
for the corresponding limiting temperature. For $D=4$,
this coincides with eq. (\ref{thagedorn}) from \cite{K-T} and agrees within 20 \% with the Unruh temperature (\ref{Tunruh})
determined by the lowest string excitation alone ($c_0=1$ in 
eq.\ (\ref{T-H})).

\medskip

Prior to the dual resonance model, Hagedorn had determined the level
density $\rho(M)$ of fireballs composed of fireballs, requiring the
same composition pattern at each level \cite{Hagedorn}. 
The resulting bootstrap condition leads to \cite{Nahm}
\be
a= 3 ~~~~~~b = r_0~\!\left[ {3 \pi \over 4} (2 \ln 2 -1)\right]^{-1/3} 
\!\simeq r_0,
\ee
where $r_0$ measures the range of the strong interaction. With $r_0 \simeq 1$
fm, we thus get $T_H \simeq 0.2$ GeV for the limiting temperature of hadronic
matter. If we identify $r_0$ with the pair production separation $x_q$
obtained in eq.\ (\ref{x-H}), we get 
\be
T_H = {1\over x_q} \simeq \sqrt{\sigma \over 2 \pi}
\ee 
and hence again agreement with the hadronic Unruh temperature 
(\ref{Tunruh}).

\medskip

Hadronic matter as an ideal gas of constituents with self-similar
composition spectra (``resonances of resonances'' or ``fireballs of
fireballs'') thus leads to an upper limit of the temperature, because
the level density of such constituents increases exponentially. What
does this have to do with the limiting acceleration found in the
$\q$ cascade of $e^+e^-$ annihilation?

\medskip

To address this problem, it is useful to recall the underlying reason
for the exponential increase of the level density in the dual resonance
model and the bootstrap model. The common origin in both cases is a
classical partition problem, which in its simplest form \cite{Blan} asks:
how many ways $\rho(M)$ are there to partition a given integer $M$ into 
ordered combinations of integers? As example, we have for $M=4$ the partitions
4, 3+1, 1+3, 2+2, 2+1+1, 1+2+1, 1+1+2, 1+1+1+1; thus here 
$\rho(M=4)=8=2^{M-1}$. It can be shown that this is generally valid, so that
\be
\rho(M) = {1\over 2} \exp\{M \ln 2\}.
\ee
For a ``gas of integers'', $T_0 = 1/\ln 2$ would thus become the limiting
temperature; the crucial feature in thermodynamics is the exponential
increase in the level density due to the equal {\sl a priori} weights 
given to all possible partitions.

\medskip

Returning now to the quark cascade in $e^+e^-$ annihilation, we note that
the form we have discussed above is a particular limiting case. We assumed
that the color field of the separating $\q$ excites in the first step one 
new pair from 
the vacuum; in principle, though with much smaller probability, it can also
excite two or more. The same is true at the next step, when the tunnelling
produces one further pair: here also, there can be two or more. Thus the
$e^+e^-$ cascade indeed provides a partition problem of the same kind.
What remains to be shown are the two specific features of our case: that 
the dominant decay chain is one where in each step one hadron is produced, 
which provides the constant deceleration of the primary quark and antiquark.

\medskip

\begin{figure}[h]
\centerline{\psfig{file=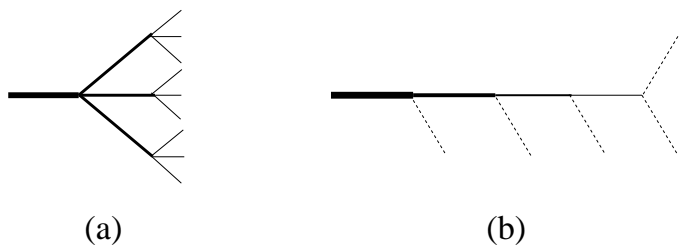,width=9cm}} 
\caption{Fireball decay patterns}
\label{branch}
\end{figure}

The statistical bootstrap model as well as the dual resonance model lead
to self-similar decay cascades, starting from a massive fireball (or
resonance), which decays into further fireballs, and so on, until at
the end one has light hadrons. In Fig.\ \ref{branch}a we illustrate 
such a cascade for the case where the average number $\bar k$ of constituents
per step in the decay (or composition) partition pattern 
\be
M \to M_{11}+M_{12}+...+M_{1k};~~~M_{11} \to M_{21}+ M_{22}+...+M_{2k};
~~~...
\ee
is $k=3$. In the statistical bootstrap model, $\bar k$ can be can be determined
\cite{Frautschi}; it is found that the crucial feature here is the power 
term multiplying the exponential increase in eq.\ (\ref{expo}). For $a < 5/2$,
the distribution in $k$ is given by
\be
F(k) = {(\ln 2)^{k-1} \over (k-1)!},
\ee
so that the average becomes
\be
\bar k = 1+2\ln 2 \simeq 2.4.
\ee
The dominant decay ($\sim 70\%$) is thus into two constituents, with 24 \%
three-body and 6 \% four-body decays. While in general the fireball mass 
$M$ could decrease in each step by $M/k$, i.e., by an amount depending on 
$M$, the case $a < 5/2$ is found to be dominated by one heavy and one
soft light hadron,
\be
M \to M_1 + h_1;~~~M_1 \to M_2+h_2;~~~...
\ee
where $h_i$ denotes final hadrons; the pattern is shown in Fig.\ 
\ref{branch}b. Moreover, the three- and four-body decays also lead to
one heavy state plus soft light hadrons. The decay thus provides a uniform 
decrease of the fireball mass by the average hadron mass or transverse 
energy. 

\medskip

We therefore conclude that the hadronization pattern we had obtained 
for $e^+e^-$ annihilation is indeed also connected to the same partition 
problem as the one leading to exponential level densities.

\vskip0.5cm

\section{Stochastic vs.\ Kinetic Thermalization}

In statistical mechanics, a basic topic is the evolution of a system of 
many degrees of freedom from non-equilibrium to equilibrium. Starting
from a non-equilibrium initial state of low entropy, the system is 
assumed to evolve as a function of time through collisions to a 
time-independent equlibrium state of maximum entropy. In other words,
the system loses the information about its initial state through a
sequence of collisions and thus becomes thermalized. In this sense,
thermalization in heavy ion collisions was studied as the transition from
an initial state of two colliding beams of ``parallel'' partons to a final 
state in which these partons have locally isostropic distributions.
This ``kinetic'' thermalization requires a sufficient density of constituents, 
sufficiently large interaction cross sections, and a certain amount
of time.     

\medskip

From such a point of view, the observation of thermal hadron production 
in high energy collisions, in particular in $e^+e^-$ and $pp$ interactions, 
is a puzzle: how could these systems ever ``have reached'' thermalization? 
Already Hagedorn \cite{Hage-born} had therefore concluded that the emitted 
hadrons were ``born in equilibrium''. Given an exponentially increasing
resonance mass spectrum, it remained unclear why collisions should 
result in a thermal system. 

\medskip

Hawking radiation provides a stochastic rather than kinetic approach to
equilibrium, with a randomization essentially provided by the quantum
physics of the Einstein-Podolsky-Rosen effect.
The barrier to information transfer due the event
horizon requires that the resulting radiation states excited from the
vacuum are distributed according to maximum entropy, with a temperature 
determined by the strength of the ``confining'' field. The ensemble of
all produced hadrons, averaged over all events, then leads to the same
equilibrium distribution as obtained in hadronic matter by kinetic
equilibration. In the case of a very high energy collision with a high 
average multiplicity already one event can provide such equilibrium;
because of the interruption of information transfer at each of the
successive quantum color horizons, there is no phase relation between
two successive production steps in a given event. The destruction of
memory, which in kinetic equilibration is achieved through sufficiently 
many successive collisions, is here automatically provided by the
tunnelling process. 

\medskip

So the thermal hadronic final state in high energy collisions is not
reached through a kinetic process; it is rather provided by successively 
throwing dice.

\vskip0.4cm

\section{Conclusions}

We have shown that quantum tunnelling through the color confinement
horizon leads to thermal hadron production in the form of Hawking-Unruh
radiation. In particular, this implies:
\begin{itemize}
\item{The radiation temperature $T_Q$ is determined by the transverse 
extension of the color flux tube, giving
\be
T_Q \simeq \sqrt{\sigma \over 2 \pi},
\ee
in terms of the string tension $\sigma$.}  
\item{The multiplicity $\nu(s)$ of the produced hadrons is 
approximately given
by the increase of the flux tube thickness with string length, leading to  
\be
\nu(s) \simeq \ln \sqrt s,
\ee
where $\sqrt s$ denotes the cms collision energy. Parton evolution and
gluon saturation will, however, increase this, as will early hard 
production. The universality of the resulting abundances is, however,
not affected.}
\item{The temperature of Hawking radiation can in general depend on the 
charge and the angular momentum of the emitting system. The former here 
provides a baryon-number dependence of the hadronization temperature 
and predicts a decrease of $T_Q$ for sufficiently high baryon density.  
The latter provides the basis for the possibility of elliptic flow 
and of a dependence of $T_Q$ on the centrality of $AA$ collisions.} 
\item{The limiting temperature obtained in the statistical bootstrap and
the dual resonance or string model arises from a self-similar composition
pattern leading to an exponentially growing level density. We find that
the underlying partition problem also leads to the cascade form obtained
for hadron emission in high energy collisions, so that the dynamic and
the thermodynamic limits have the same origin.}
\item{In statistical QCD, thermal equilibrium is reached kinetically
from an initial non-equilibrium state, with memory destruction through
successive interactions of the constituents. In high energy collisions, 
tunnelling prohibits information transfer and hence leads to stochastic 
production, so that we have a thermal distribution from the outset.} 
\end{itemize}

\medskip

We close with a general comment. In astrophysics, Hawking-Unruh radiation
has so far never been observed. The thermal hadron spectra in high 
energy collisions may thus indeed be the first experimental instance
of such radiation, though in strong interaction instead of gravitation.

\vskip1cm

\centerline{\bf \Large Acknowledgements}

\bigskip

D.\ K.\ is grateful to E.\ Levin and K.\ Tuchin for useful and stimulating
discussions. H.\ S.\ thanks M.\ Kozlov for helpful discussions and a critical 
reading to the text. The work of D.\ K.\ was supported by the U.S. 
Department of Energy under Contract No. DE-AC02-98CH10886.


\vskip1cm

\begin{thebibliography}{99}

\bibitem{C-R} E.\ Recami and P.\ Castorina, Lett. Nuovo Cim. 15 (1976) 347.

\bibitem{Salam} A.\ Salam and J.\ Strathdee, \PR D18 (1978) 4596;\\ 
see also C.J.\ Isham, A.\ Salam, and J.\ Strathdee, \PR D3 (1971) 867. 

\bibitem{Gross:1973id} D.~J.~Gross and F.~Wilczek,
 Phys.\ Rev.\ Lett.\  {\bf 30}, 1343 (1973); \\
 H.~D.~Politzer, Phys.\ Rev.\ Lett.\  {\bf 30}, 1346 (1973).

\bibitem{scale1}
 J.~R.~Ellis, Nucl.\ Phys.\  B {\bf 22}, 478 (1970);\\
 R.~J.~Crewther,  Phys.\ Lett.\  B {\bf 33}, 305 (1970); \\
 M.~S.~Chanowitz and J.~R.~Ellis, Phys.\ Lett.\  B {\bf 40}, 397 (1972);\\
 J.~Schechter, Phys.\ Rev.\  D {\bf 21}, 3393 (1980).
\bibitem{scale2}
J.~C.~Collins, A.~Duncan and S.~D.~Joglekar,
Phys.\ Rev.\  D {\bf 16}, 438 (1977);\\
N.~K.~Nielsen, Nucl.\ Phys.\  B {\bf 120}, 212 (1977).
  
\bibitem{MSh} A.\ A.\ Migdal and M.\ A.\ Shifman, \PL B 114 (1982) 445.

\bibitem{KLT-Pom} 
D.~Kharzeev, E.~Levin and K.~Tuchin, Phys.\ Lett.\ B {\bf 547}, 21 (2002);
Phys.\ Rev.\ D {\bf 70}, 054005 (2004)

\bibitem{SVZ} M.\ A.\ Shifman, A.\ I.\ Vainshtein and V.\ I.\ Zakharov,
\NP B 147 (1979) 385.

\bibitem{Maldacena:1997re}
J.~M.~Maldacena, Adv.\ Theor.\ Math.\ Phys.\  {\bf 2}, 231 (1998)
[Int.\ J.\ Theor.\ Phys.\  {\bf 38}, 1113 (1999)]

\bibitem{Hawking} S.\ W.\ Hawking, Comm.\ Math.\ Phys.\ 43 (1975) 199. 

\bibitem{Unruh} W.\ G.\ Unruh, \PR D14 (1976) 870.

\bibitem{Grillo} A.\ F.\ Grillo and Y.\ Srivastava, \PL B 85 (1979) 377.

\bibitem{Barshay} S.\ Barshay and W.\ Troost, \PL 73B (1978) 437

\bibitem{Hosoya} A.\ Hosoya, Progr.\ Theoret.\ Phys.\ 61 (1979) 280.

\bibitem{Horibe} M.\ Horibe, Progr.\ Theoret.\ Phys.\ 61 (1979) 661.
 
\bibitem{K-T} 
D.~Kharzeev and K.~Tuchin, Nucl.\ Phys.\  A {\bf 753}, 316 (2005);\\
D.~Kharzeev, Nucl.\ Phys.\  A {\bf 774}, 315 (2006)

\bibitem{KLT} 
  D.~Kharzeev, E.~Levin and K.~Tuchin,
  Phys.\ Rev.\  C {\bf 75}, 044903 (2007).
  
\bibitem{Hagedorn} R.\ Hagedorn, Nuovo Cim.\ Suppl.\ 3 (1965) 147;
Nuovo Cim.\ A 56 (1968) 1027  

\bibitem{species} F.\ Becattini, \ZP C69 (1996) 485 ($e^+e^-$);\\
F.\ Becattini and U.\ Heinz, \ZP C76 (1997) 268 ($pp/p\bar p$);\\
J.\ Cleymans and H.\ Satz, \ZP C57 (1993) 135 (heavy ions);\\
F.\ Becattini et al., \PR C64 (2001) 024901 (heavy ions);\\
P.\ Braun-Munziger, K.\ Redlich and J.\ Stachel, (heavy ions).

\bibitem{T-c} See e.g., M.\ Cheng et al., \PR D 74 (2006) 054507,
for the latest state and references to earlier work. 

\bibitem{TD} T.\ D.\ Lee, \NP B264 (1986) 437.

\bibitem{P-W} M.\ K.\ Parikh and F.\ Wilczek, \PRL 85 (2000) 5042.

\bibitem{DdD} J.\ Dias de Deus and C.\ Pajares, hep-ph/0605148

\bibitem{Ruf} See e.g., Li Zhi Fang and R.\ Ruffini, {\sl Basic Concepts in 
Relativistic Astrophysics}, World Scientific, Singapore 1983.

\bibitem{Bekenstein} J.\ D.\ Bekenstein, \PR D 7 (1973) 2333.

\bibitem{alpha} R.\ Alkofer, C.\ S.\ Fischer and F.\ J.\ Llanes-Estrada,
\PL B6111 (2005) 279.

\bibitem{Novello} M.\ Novello et al., \PR D61 (2000) 045001.

\bibitem{TD-vac} T.\ D.\ Lee, in {\sl Statistical Mechanics of Quarks and
Hadrons}, H.\ Satz (Ed.), North Holland Publishing Co., Amsterdam 1980.


\bibitem{L-QCD} H.\ Pagels and E.\ T.\
 Tomboulis, \NP B143 (1978) 453;\\
P.\ Castorina and M.\ Consoli, \PR D 35 (1987) 3249;\\
L.\ B.\ Abbott, \NP B185 (1981) 189;\\
L.\ Maiani et al., \NP B273 (1986) 275.

\bibitem{Raufeisen} D.\ Kharzeev and J.\ Raufeisen, nucl-th/0206073.

\bibitem{Gerlach} U.\ H.\ Gerlach, J.\ Mod.\ Phys.\ A 11 (1996) 3667.

\bibitem{Terashima} H.\ Terashima, \PR D 61 (2000) 104016.

\bibitem{EPR} A.\ Einstein, B.\ Podolsky and N.\ Rosen, \PR 47 (1935) 777;\\
Y.\ Aharanov and D.\ Bohm, \PR 108 (1957) 1070.

\bibitem{Pauli} For a clear discussion and references to the original
solutions by M.\ Born (1909) and A.\ Sommerfeld (1910),, see 
W.\ Pauli, {\sl Relativit\"atstheorie}, in {\sl Enzyklop\"adie der 
ma\-the\-matischen Wissenschaften}, Teubner Verlag, Leipzig 1921; English
version {\sl Theory of Relativity}, Pergamon Press, 1958.

\bibitem{Laflamme} R.\ Laflamme, \PL B 196 (1987) 449.

\bibitem{Schwinger} J.\ Schwinger, \PR 82 (1951) 664.

\bibitem{bj} J.\ D.\ Bjorken, Lecture Notes in Physics (Springer) 56 
(1976) 93.

\bibitem{nus} A.\ Casher, H.\ Neuberger and S.\ Nussinov, \PR D20 (1979) 179.

\bibitem{I-U-W} S.\ Iso, H.\ Umetsu and F.\ Wilczek, hep-th/0606018 

\bibitem{Blan} Ph.\ Blanchard, S.\ Fortunato and H.\ Satz, \EP 34 (2004) 361. 

\bibitem{Luescher} M.\ L\"uscher, G.\ M\"unster and P.\ Weisz, \NP B180
(1981) 1.

\bibitem{Gribov} V.~N.~Gribov, arXiv:hep-ph/0006158.
    
\bibitem{Bali} G.\ Bali, K.\ Schilling and C.\ Schlichter, \PR D 51 (1995)
5165. 

\bibitem{BCKSR} F.\ Becattini et al., \PR C 64 (2001) 024901

\bibitem{Sikler} See e.g., F.\ Sikler et al.\ (NA49), \NP A661 (1999) 45c.

\bibitem{drm}  S.~Fubini and G.~Veneziano, Nuovo Cim. 64A (1969) 811;\\
K.~Bardakci and S.~Mandelstam, Phys. Rev. 184 (1969) 1640.

\bibitem{h-w} K.\ Huang and S.\ Weinberg, \PRL 25 (1970) 895.

\bibitem{Nahm} W.\ Nahm, \NP 45 B (1972) 525.

\bibitem{Frautschi} S.\ Frautschi, \PR D3 (1971) 2821.

\bibitem{Hage-born} R.\ Hagedorn, {\sl Thermodynamics of Strong Interactions},
CERN 71-12, 1971.

\end{thebibliography}
\end{document}